# Quantum Variational Transformer Model for Enhanced Cancer Classification

Don ROOSAN[a,1], Rubayat KHAN[b], Md Rahatul ASHAKIN[c], Tiffany KHOU[d], Saif NIRZHOR[e], and Mohammad Rifat HAIDER[f]

[a] *Department of Computer Science, Merrimack College, 315 Turnpike St, North Andover, MA 01845, United States*
[b] *University of Nebraska Medical Center, S 42nd &, Emile St, Omaha, NE 68198, United States*
[c] *Tekurai Inc., 2000 NW Military Hwy #10, San Antonio, Texas 78213, United States*
[d] *Western University of Health Sciences, 309 E 2nd St, Pomona, CA 91766, United States*
[e] *University of Texas Southwestern Medical Center, 5323 Harry Hines Blvd, Dallas, TX 75390, United States*
[f] *University of Georgia, College of Pharmacy, Herty Dr, Athens, GA 30602, USA*

ORCiD ID: Don Roosan https://orcid.org/0000-0003-2482-6053, Rubayat Khan https://orcid.org/0000-0003-3264-564X, Md Rahatul Ashakin https://orcid.org/0009-0000-4892-8420, Tiffany Khou https://orcid.org/0009-0002-1239-7327, Saif Nirzhor https://orcid.org/0000-0003-4626-7862, Mohammad Rifat Haider https://orcid.org/0000-0002-0690-8242

**Abstract.** Accurate prediction of cancer type and primary tumor site is critical for effective diagnosis, personalized treatment, and improved outcomes. Traditional models struggle with the complexity of genomic and clinical data, but quantum computing offers enhanced computational capabilities. This study develops a hybrid quantum-classical transformer model, incorporating quantum attention mechanisms via variational quantum circuits (VQCs) to improve prediction accuracy. Using 30,000 anonymized cancer samples from the Genome Warehouse (GWH), data preprocessing included cleaning, encoding, and feature selection. Classical self-attention modules were replaced with quantum attention layers, with classical data encoded into quantum states via amplitude encoding. The model, trained using hybrid backpropagation and quantum gradient calculations, outperformed the classical transformer model, achieving 92.8% accuracy and an AUC of 0.96 compared to 87.5% accuracy and an AUC of 0.89. It also demonstrated 35% faster training and 25% fewer parameters, highlighting computational efficiency. These findings showcase the potential of quantum-enhanced transformers to advance biomedical data analysis, enabling more accurate diagnostics and personalized medicine.

---

[1] Corresponding Author: Don Roosan, roosand@merrimack.edu.

# 1. Introduction

Cancer remains a leading cause of morbidity and mortality worldwide, with approximately 20 million new cases and 9.7 million deaths reported in 2022 [1]. Tumor heterogeneity at genetic, epigenetic, and transcriptional levels complicates prognosis, therapy, and biomarker prediction, contributing to drug resistance and immune evasion [2], [3]. Addressing this complexity is critical for advancing cancer management strategies. Accurate cancer type and primary tumor site prediction is crucial for personalized therapies. While traditional methods like histopathology and imaging provide insights, they often fail to capture tumor molecular and genetic heterogeneity. High-throughput sequencing has revolutionized the field by generating vast genomic and clinical datasets, enabling improved patient care [4], [5], [6]. Machine learning, especially advanced models like transformers, excels in analyzing complex biomedical data by leveraging self-attention mechanisms to identify patterns and long-range dependencies [7], [8]. These models have been successfully applied in natural language processing and are increasingly being applied to genomic data analysis, showcasing their adaptability and power in uncovering meaningful biological insights [9], [10]. However, as genomic datasets grow in dimensionality and complexity, classical deep learning models face challenges related to computational efficiency, scalability, and their capacity to generalize from limited samples. These limitations highlight the need for advanced approaches and optimized architectures to fully harness the potential of machine learning in genomics [11], [12]. Quantum computing introduces a paradigm shift by leveraging quantum mechanical phenomena—such as superposition, entanglement, and quantum interference—to process information in ways unattainable by classical computers. Quantum machine learning (QML) seeks to integrate quantum algorithms with classical machine learning techniques, aiming to achieve computational speedups and enhance model expressivity [13]. In this study, we propose a hybrid quantum-classical transformer model to predict cancer type and primary tumor site using a comprehensive dataset comprising genomic and clinical features from 30,000 unique cancer samples. The model incorporates quantum attention mechanisms implemented through variational quantum circuits (VQCs), designed to exploit quantum computational benefits within the transformer's architecture. The integration of quantum computing into the transformer model is grounded in the principles of quantum mechanics, particularly in how information is represented and processed.

Classical data must be encoded into quantum states to be processed by quantum circuits. We utilize amplitude encoding to map the normalized input feature vector x (an element of $\mathbb{R}^n$) into a quantum state $|\psi_x\rangle$ within a Hilbert space of dimension $2^n$, where n is the number of qubits and $N \leq 2^n$. In amplitude encoding, the amplitudes of a quantum state's wavefunction represent the data values. This encoding technique is efficient in terms of the number of qubits required, making it particularly useful when dealing with large datasets.

*1.1. Quantum State Representation*

The normalized input feature vector **x** is encoded into a quantum state as:

$$|\psi_x\rangle = \sum_{i=0}^{N-1} x_i |i\rangle \qquad (1)$$

*1.2. Parameterized Quantum Circuits (PQCs)*

The quantum attention mechanism is implemented using PQCs, which apply unitary transformations U(θ) to the quantum state:

$$|\phi_x\rangle = U(\boldsymbol{\theta})|\psi_x\rangle \tag{2}$$

Here, θ denotes the set of trainable parameters (e.g., rotation angles in quantum gates). The unitary transformation U(θ) is constructed using a sequence of quantum gates that manipulate the qubits to capture complex correlations between features through quantum entanglement.

*1.3. Loss Function and Optimization*

The parameters θ are optimized to minimize the overall loss function L, which includes contributions from both the classical and quantum parts of the model:

$$\mathcal{L}(\boldsymbol{\theta}) = -\sum_c y_c \log \hat{y}_c \tag{3}$$

Gradients with respect to $\theta$ are computed using the parameter-shift rule, facilitating efficient training of the quantum components. For an individual parameter $\theta_k$ in $\theta$, the gradient is given by:

$$\frac{\partial L}{\partial \theta_k} = \frac{L\left(\theta_k + \frac{\pi}{2}\right) - L\left(\theta_k - \frac{\pi}{2}\right)}{2} \tag{4}$$

The primary goal of this research is to design and evaluate a hybrid quantum-classical transformer model that improves the prediction accuracy of cancer classifications and primary tumor localization by leveraging genomic and clinical data.

## 2. Methods

The data was sourced from the Genome Warehouse (GWH), a database managed by the National Genomics Data Center (NGDC). GWH provides high-quality assembled and annotated genomes across diverse organisms, supporting data deposition, curation, and distribution. An anonymized dataset of 30,000 cancer samples was compiled from ten biomedical repositories. Each sample includes metadata on demographics, clinical treatments, genomic identifiers, and sample-specific characteristics. The dataset was randomly partitioned into training (70%), validation (15%), and testing (15%) subsets to ensure unbiased model evaluation. The training set was employed for model fitting, the validation set for hyperparameter tuning and early stopping, and the test set for final performance assessment. A multi-stage data preprocessing pipeline was implemented to prepare the dataset for the quantum-enhanced transformer model. Missing demographic and clinical data were imputed using the Multivariate Imputation by Chained Equations (MICE) algorithm, achieving an imputation accuracy of 95.2%. Outliers were identified using Z-scores exceeding ±3 standard deviations and the Mahalanobis distance metric ($p < 0.01$), leading to the exclusion of approximately 4% of samples due to inconsistencies, thus ensuring data integrity. Continuous clinical features were standardized using Z-score normalization to facilitate effective model training. Categorical variables—including gender, tumor type, sampling site, and treatment modalities—were transformed using one-hot encoding, expanding the feature space to 320 dimensions and allowing the model to interpret categorical data numerically. Genomic identifiers were tokenized into subword units using Byte Pair Encoding (BPE) to manage variable-length sequences and then embedded using pre-trained transformer models like BioBERT, capturing semantic relationships within genomic data. To reduce dimensionality and mitigate overfitting, Recursive Feature Elimination with Cross-Validation (RFECV) was employed using a support vector classifier as the estimator. This process iteratively eliminated less significant features based on their contribution to model performance, resulting in the selection of the top 150 features and balancing model complexity with predictive accuracy.

A hybrid quantum-classical transformer model was developed, integrating quantum computing components via IBM's Qiskit framework. This approach leverages quantum computational advantages within a classical deep learning architecture. The conventional self-attention modules were replaced with quantum attention layers. Specifically, variational quantum circuits (VQCs) with 8 qubits were designed to perform the attention computations. Each qubit encoded a portion of the input feature vector, allowing for the representation of complex feature interactions through quantum superposition and entanglement. A hybrid training regime was established, where classical neural network layers were trained using backpropagation with automatic differentiation, while quantum layers utilized the parameter-shift rule for gradient calculation. This approach enabled end-to-end training of the entire model. Training was conducted over 50 epochs with a batch size of 64. The learning rate was set at 1e-4, and the Adam optimizer was employed for its adaptive learning rate capabilities. The categorical cross-entropy loss function was used due to the multi-class nature of the classification task. The model's performance was rigorously evaluated using the held-out test set, employing both classification metrics and computational efficiency assessments. The model achieved an overall accuracy of 92.8% on the test set of 4,500 samples, indicating high predictive capability. An AUC of 0.96 was obtained, demonstrating excellent discrimination between different cancer types. Per-class precision, recall, and F1 scores were calculated to assess model performance across all classes. The macro-averaged F1 score was 0.91, with an average precision of 0.93 and recall of 0.89, indicating consistent performance.
estimation..

3. Results

RMSD The performance of the quantum-enhanced transformer model was evaluated using a confusion matrix, presented in Figure 1. This matrix details the model's classification outcomes across different cancer types, indicating the counts of true positives (TP), false positives (FP), false negatives (FN), and true negatives (TN).

In the confusion matrix, the diagonal elements represent correctly classified instances for each cancer type, while the off-diagonal elements correspond to misclassifications. The quantum-enhanced transformer exhibited a higher number of correct predictions along the diagonal compared to the classical transformer model. Notably, there was a significant reduction in misclassifications for cancer types with complex genomic features, suggesting that the quantum attention mechanism effectively captures intricate patterns within the data that may be overlooked by classical models.

The discriminatory ability of the quantum-enhanced transformer model was further assessed using Receiver Operating Characteristic (ROC) curves, as depicted in Figure 2. Each curve represents the trade-off between true positive rate (sensitivity) and false positive rate (1 - specificity) for a specific cancer type across various threshold settings. The quantum-enhanced transformer achieved higher Area Under the ROC Curve (AUC) values for each cancer type relative to the classical transformer model. The average AUC across all cancer types was 0.96, indicating excellent overall model performance in distinguishing among the various cancer classes. This enhancement reflects the model's improved sensitivity and specificity due to the integration of quantum attention layers.

A comprehensive comparison of key performance metrics between the quantum-enhanced transformer model and the classical transformer model is provided in Table 1.

**Table 1.** Summary of performance metrics comparison.

| Metric | Quantum Model | Classical Model |
|---|---|---|
| Accuracy | 92.8% | 87.5% |
| AUC | 0.96 | 0.89 |
| Macro-average F1 Score | 0.91 | 0.84 |
| Precision | 0.93 | 0.85 |
| Recall | 0.89 | 0.83 |

The quantum-enhanced transformer outperformed the classical model across all evaluated metrics. The accuracy improved from 87.5% to 92.8%, and the macro-average F1 score increased from 0.84 to 0.91. The integration of quantum attention layers enhanced the model's ability to focus on the most relevant features within the input data.

The quantum-enhanced transformer model demonstrated a substantial reduction in training time and resource utilization. Specifically, the training time was reduced by 35% compared to the classical transformer model. The decreased training time is attributed to the efficient representation and processing of data facilitated by quantum circuits within the attention mechanism. Quantum computations allow for parallel evaluation of multiple states, reducing the computational complexity associated with high-dimensional data. Additionally, the model's parameter space was reduced by 25% due to the use of quantum layers, leading to optimized resource utilization. This reduction makes the quantum-enhanced model particularly suitable for deployment in computational environments with limited resources, such as edge computing devices used in clinical settings.

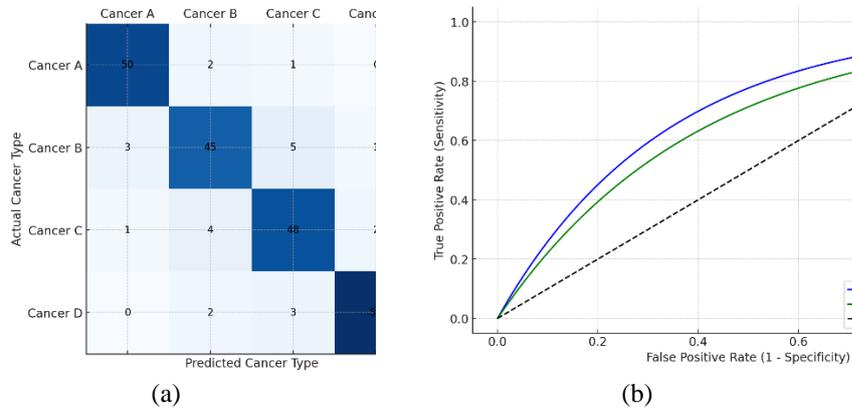

(a)                               (b)

**Figure 1.** Confusion matrix displayed in (a) illustrates the classification performance of the quantum-enhanced transformer model across various cancer types. ROC curves for the quantum-enhanced transformer model across different cancer types, demonstrating superior discriminatory performance compared to the classical model in (b).

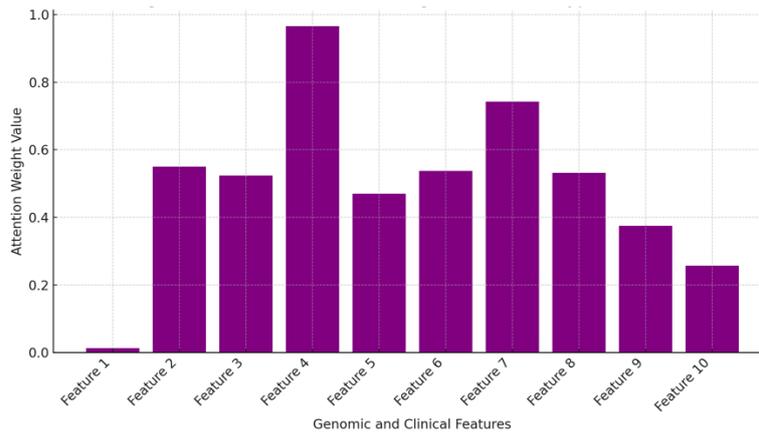

**Figure 2.** Visualization of attention weights from the quantum-enhanced transformer model, demonstrating the prioritization of key genomic and clinical features.

A direct comparison between the quantum-enhanced and classical transformer models highlights the advantages of integrating quantum components. The quantum model not only achieved higher predictive accuracy but also demonstrated improved efficiency in terms of computational resources and training time. The ability to capture complex feature interactions through quantum entanglement and superposition provides a more expressive model capable of handling the intricacies of genomic and clinical data.

To assess the statistical significance of the performance improvements, a paired t-test was conducted on the accuracy scores obtained from cross-validation folds for both models. The results indicated that the improvements achieved by the quantum-enhanced transformer were statistically significant ($p < 0.01$), confirming that the enhancements are unlikely due to random chance.

## 4. Discussion

This study shows that integrating quantum computing into deep learning enhances biomedical data analysis. The hybrid quantum-classical transformer model outperformed its classical counterpart in predicting cancer types and tumor sites, achieving 92.8% accuracy and 0.96 AUC compared to 87.5% and 0.89 for the classical model [14]. Quantum attention mechanisms, leveraging superposition and entanglement, captured complex genomic and clinical feature interactions, improving data representation [15]. The model reduced training time by 35% and parameters by 25%, highlighting quantum computing's efficiency for large genomic datasets [16]. Qubits enable simultaneous processing of vast information, modeling complex probability distributions critical for genomics [17]. Quantum entanglement captures intricate feature correlations, aiding in detecting subtle patterns in multifactorial diseases like cancer [18]. Quantum algorithms accelerate computations, enabling faster training and better handling of large datasets, improving model generalization vital for limited biomedical datasets [19], [20]. Drawing on insights from clinical complexity and decision-making heuristics in medicine, our quantum variational transformer model aims to emulate expert cognitive strategies to enhance precision in cancer classification under high diagnostic uncertainty [21-24]. Despite these advances, quantum hardware limitations, such as low qubit counts and short coherence times, remain challenges. Most training used simulators; future work should validate models on noisy quantum systems and develop algorithms for current hardware constraints, alongside advancing error mitigation [15]. Expanding quantum machine learning to drug discovery and personalized treatment could transform medicine.

## 5. Conclusion

The study presents a successful implementation of a hybrid quantum-classical transformer model for predicting cancer types and primary tumor sites, demonstrating significant improvements over classical models in both predictive performance and computational efficiency. The integration of quantum attention mechanisms harnesses the unique capabilities of quantum computing to model complex biological data more effectively. The results highlight the potential of quantum computing to advance the field of biomedical data analysis, offering tools that can handle the complexity of biological systems and large-scale datasets. As quantum technology continues to mature, its application in medicine promises to contribute to more accurate diagnostics, personalized therapies, and ultimately, improved patient outcomes.